\UseRawInputEncoding
\documentclass[preprint,onecolumn,nofootinbib]{revtex4}
\pdfoutput=1
\usepackage[colorlinks=true,linkcolor=blue,urlcolor=blue,filecolor=black,citecolor=red,pdfstartview=FitV,pdftitle={},pdfsubject={},pdfkeywords={},pdfpagemode=None,bookmarksopen=true]{hyperref}
\usepackage{graphicx}%Include figure files
\usepackage{amsmath}
\usepackage{amsfonts}
\usepackage{amssymb,ulem}
\usepackage{color,xcolor}%
\usepackage{CJK}
\usepackage{subfigure}

\usepackage{dcolumn}% Align table columns on decimal pointl
\usepackage{MnSymbol,wasysym}
\usepackage{braket}

\usepackage{eurosym}
\usepackage{calrsfs}

\begin{document}
\title{Complexity growth of BTZ black hole in massive gravity with a null string}

\author{Yu-Ting Zhou$^{1,}$$^{2}$}
\email{yu-tingzhou@nuaa.edu.cn}
%\author{xxx $^{1}$,$^{2}$}
%\email{xxxxxxxxx@xx.xx.xx}
\affiliation{
  $^1$ College of Physics, Nanjing University of Aeronautics and Astronautics, Nanjing 210016, China\\
  $^2$ Key Laboratory of Aerospace Information Materials and Physics (NUAA), MIIT, Nanjing 211106, China.}

\begin{abstract}
In this paper, we investigate the complexity growth of the tensionless limit of string in the neutral BTZ black hole horizon in massive gravity. When the string approaches the horizon, we observe a novel phenomenon for the Nambu-Goto action growth that produces significant difference from tensile string geometry. The string's tension is then suggested to partially contribute to the growth of the action. We also argue a potential proposal that reconstructs the complexity from the {renormalization} group (RG) flow. 
\end{abstract}

\maketitle
\tableofcontents

\section{Introduction}
The nature of gravity or  spacetime is one of important topics in modern theoretical physics. To disclose their nature,  one potential way is to study the black holes, which provide an arena for general relativity and quantum field theory. Inspired by the black hole thermodynamics, physicists realized that the gravity is holographic and constructed the gauge/gravity duality, and one of its implement is the AdS/CFT correspondence \cite{Maldacena:1997re,Gubser:1998bc,Witten:1998qj} which  connects a d-dimensional gravity theory in AdS spacetime and a d-1 dimensional  conformal field theories (CFT) without gravity. The applications of the duality allows us to investigate gravity and strongly coupled system in a novel perspective.

In the framework of holography, the entanglement entropy (EE) which measures the correlation among the strongly coupled subsystems in a field theory, has an elegant geometric description in the dual bulk theory. The statement is that EE between the subregion on the dual boundary is proportional to the minimal surface in the bulk geometry, which is now called the Hubey-Rangamani-Takayanagi (HRT) surface \cite{Takayanagi:2012kg,Hubeny:2007xt}. Later, Susskind et al. suggested that the entanglement is not the whole story because the EE cannot provide complete information about the interior black hole. Thus, they proposed using computational complexity to calculate the size of the wormhole that connected the two sides of the AdS black hole's boundary times \cite{Stanford:2014jda,Susskind:2014moa,Susskind:2014jwa,Susskind:2014rva,Susskind:2018pmk}. The concept of complexity originates from quantum information theory \cite{Osborne_2012,TCS-066,Dvali000,PhysRevA.94.040302,PhysRevD.96.126001,Watrous:2008,Bao:2018ira}, which quantifies the difficulty of changing from one quantum state to another. The problem here with complexity is that it is difficult to  clearly and precisely define the initial states and target states in quantum field theory. Despite numerous efforts in this direction have been made \cite{Vanchurin:2016met,Chapman:2017rqy,Molina-Vilaplana:2018sfn,Bhattacharyya:2018wym,Nielsen:2005,Nielsen:2006,Jefferson:2017sdb,Yang:2018nda,Bhattacharyya:2018bbv,Bhattacharyya:2019kvj,Camargo:2022wkd,Adhikari:2022whf,Adhikari:2021pvv,Adhikari:2022oxr}, a well-defined complexity remains an open question. In holographic aspect, two geometric proposals or conjectures have been suggested to describe complexity. One is the `` complexity=volume (CV) " conjecture where  $V$ is the volume of the Einstein-Rosen (ER) bridge connecting the two sides of the AdS black hole's boundary times.  The other proposal is the `` complexity=action (CA) " conjecture. Here the $A$ is the classical action of a space-time region that is enclosed by the bulk Cauchy slice anchored at the boundaries, and it was also known as the ``Wheeler-Dewitt (WdW)" patch \cite{Chapman:2016hwi,Brown:2015bva,Brown:2015lvg}.

There have been plenty of works on the application of the CA conjecture in stationary systems \cite{Pan:2016ecg,Guo:2017rul,Momeni:2016ekm, Tao:2017fsy,Alishahiha:2017hwg,Reynolds:2017lwq,Qaemmaqami:2017lzs,Sebastiani:2017rxr,Couch:2017yil,Swingle:2017zcd,Cano:2018aqi,Chapman:2018dem,Chapman:2018lsv,Auzzi:2018pbc,Yaraie:2018hwz,Alishahiha:2018tep,An:2018xhv,Cai:2016xho,Ghodsi:2020qqb,Frassino:2019fgr} and references therein. Parallel works can also be seen in dynamic systems,  including the complexity growth with probe branes \cite{Abad:2017cgl} and non-local operator studies in the BTZ black hole \cite{Ageev:2014nva}. In those studies,  the time-dependent process was inspired by the work of jet-quenching phenomena in ion collisions and  the system is moved by a drag force \cite{Gubser:2006bz}. This phenomenon was explained by the fact that as a charged particle passed through a quark-gluon plasma, it would lose energy because of the effect of shear viscosity. According to AdS/CFT correspondence, these could be viewed as an open string with one end point attached to the boundary and the other end point approaching the horizon, accompany with  a  momentum flow along the string. Thus, the most useful method to analyze this dissipative system is to use a probe. More recently, Nagasaki investigated the complexity of the $AdS_{5}$ black hole via CA conjecture by inserting a probe string that moves in a circle in the spatial part of the AdS spacetime \cite{Nagasaki:2017kqe}. These studies are soon extended to accelerating black holes \cite{Nagasaki:2021ldz, Nagasaki:2022lll}, rotating black holes \cite{Nagasaki:2018csh,Nagasaki:2019icm} and black holes in Horndeski gravity \cite{Santos:2020xox,Bravo-Gaete:2020lzs}. Furthermore, Abdulrahim Al Balushi et al. propose that thermodynamic volume has a significant effect on the complexity of the formation of black holes in both CV and CA conjectures and  show a deep connection between them \cite{AlBalushi:2020rqe,AlBalushi:2020heq}. Then in \cite{Bernamonti:2021jyu,Zhang:2022quy}, the authors study the effects of angular momentum on complexity formation and the complexity of rotating black holes with conical deficits, respectively. Later, considering that a realistic system should have momentum relaxation, we generalize the study into massive BTZ black hole  \footnote{For convenience, we will use ``massive BTZ black hole'' as the abbreviation for ``BTZ black hole in massive gravity '' in the below content.}\cite{Zhou:2021vsm}.

As we mentioned before, the previous strategy in applying CA into the dynamic systems is to insert a string at the boundary which then moves in the spatial part of the AdS spacetime. Then an interesting question is  what can one observe in the complexity growth when the string is put near the horizon? Since the complexity growth could be different between the UV and IR sides as the EE behaves \cite{Nishioka:2018khk},  this question becomes more significant when we recall the Wilsonian idea of renormalization group (RG) flow theory \cite{POLCHINSKI1984269,WILSON197475} and holographic principle. The original idea that the complexity has dynamic under an RG flow has been discussed based on the circuit complexity in \cite{Bhattacharyya:2018bbv} and \cite{Bhattacharyya:2019kvj}, in which the authors employed the circuit complexity as a probe to analyze a certain flow of effective coupling. Their proposal gives a modification to Nielsen’s circuit complexity calculation and has the same behavior as in holography via the framework of path-integral optimization. Morever, A. Banerjee et al. also used the circuit complexity as a probe of RG flow to study the different quantum field theories \cite{Banerjee:2022ime}. All those studies provide strong signals that the complexity could be viewed as a dynamic quantity. On the other hand, the near-horizon string behaves very differently from the normal string \cite{Bagchi:2021ban}. Especially, when the string approaches the horizon, its tension will vanish and produce a new geometric structures such that the original Riemannina geometry will be replaced by Carrolian geometry. In the mathematical viewpoint, the algebra structures of Carrolian geometry were isomorphic to the Bondi-Metzner-Sachs (BMS) algebras \cite{Bagchi:2010zz,Bagchi:2012cy,Duval:2014lpa,Duval:2014uva}, which were nothing but the asymptotic symmetries of Minkovski spacetime \cite{Bondi:1962, PhysRev.128.2851}, so it can emerge in asymptotically flat spacetimes in the holographic context \cite{Bagchi:2010zz,Bagchi:2012cy,Barnich:2012aw,Bagchi:2012yk,Bagchi:2012xr,Barnich:2012xq,Barnich:2012rz,Bagchi:2013lma,Bagchi:2014iea,Hartong:2015xda,Hartong:2015usd,Bagchi:2016bcd}. The Carrollian structure  has been extensively applied in physics and astrophysics, such as in cosmology \cite{deBoer:2021jej}, black hole horizon physics \cite{Donnay:2019jiz}, and the explanation of black hole entropy \cite{Carlip:2017xne,Carlip:2019dbu,Bagchi:2022iqb}.

Thus, the aim of our work is to investigate the complexity growth in massive BTZ black hole with a probe tensionless string. We shall investigate  how the complexity encodes information about the tensionless string.  The novelty of this work is that we consider the near-horizon limit of a string when it approaches the black hole horizon. Massive terms in gravity theory usually induce the ghost, which causes instability, but there have recently been many works to construct massive gravity theories that avoid the instability \cite{deRham:2010ik,deRham:2010kj,Hassan:2011hr,Hassan:2011vm,Hassan:2011tf}. In the AdS/CFT context, the massive terms in the gravitational action break the diffeomorphism symmetry in the bulk, which corresponds to momentum dissipation in the dual boundary field theory \cite{Blake:2013bqa}. It is noted that some of us considered a Wilson line operator in the massive theory, which is dual to a dynamical system where non-local operator could correspond to a particle moving on the boundary gauge theory with momentum relaxation \cite{Zhou:2021vsm}. Here the subtle difference  is that we take the string near the horizon and focus on the influence of graviton mass and black hole mass on the velocity-dependent complexity growth, which is dual to Nambu-Goto action growth in a massive black hole with a probe null string and we find some novel phenomena comparing to tensile string case.

The rest of the paper is organized as follows. In Sec. \ref{mBTZNHE}, we briefly review the massive BTZ black hole and its near-horizon expansion. In Sec. \ref{Horizon StringC}, we give the Nambu-Goto action of the near-horizon limit of a tensile string in a massive BTZ black hole and then derive the action growth of this tensionless string. Then we analyze the effect of the graviton mass, black hole mass, and string velocity on the features of Nambu-Goto action growth. We summarize our work in the last section.

\section{Massive BTZ black hole and its near-horizon expansion}\label{mBTZNHE}

We start from the 3-dimensional Einstein-massive gravity with the action \cite{Hendi:2016pvx,Chougule:2018cny}
\begin{equation}
\mathcal{I}=-\frac{1}{16\pi}\int d^{3}x \sqrt{-g}\left[\mathcal{R}-2\Lambda+m^{2}\sum_{i=0}^{4}c_{i}\mathcal{U}_{i}(g,f)\right],
\label{1}
\end{equation}
where $\mathcal{R}$ is the scalar curvature, $\Lambda=-1/L^{2}$ is the cosmological constant, and the last terms are Fierz-Pauli mass terms, with $m$ denoting the mass of graviton \cite{Creminelli:2005qk}. Here, $f$ is a fixed symmetric tensor, $c_{i}$'s are constant coefficients, and $\mathcal{U}_i$ are symmetric polynomails of the eigenvalues of the $d\times d$ matrix $\mathcal{K}^{\mu}_{\nu}=\sqrt{g^{\mu \alpha}f_{\alpha \nu}}$, which have the formulas \cite{deRham:2010kj,Hinterbichler:2011tt}
\begin{eqnarray}
\mathcal{U}_{1} &=&\left[ \mathcal{K}\right] ,\;\;\;\;\;\mathcal{U}_{2}=%
\left[ \mathcal{K}\right] ^{2}-\left[ \mathcal{K}^{2}\right] ,\;\;\;\;\;%
\mathcal{U}_{3}=\left[ \mathcal{K}\right] ^{3}-3\left[ \mathcal{K}\right] %
\left[ \mathcal{K}^{2}\right] +2\left[ \mathcal{K}^{3}\right] ,  \notag \\
&&\mathcal{U}_{4}=\left[ \mathcal{K}\right] ^{4}-6\left[ \mathcal{K}^{2}%
\right] \left[ \mathcal{K}\right] ^{2}+8\left[ \mathcal{K}^{3}\right] \left[
\mathcal{K}\right] +3\left[ \mathcal{K}^{2}\right] ^{2}-6\left[ \mathcal{K}%
^{4}\right].
\label{2}
\end{eqnarray}
By variating the action (\ref{1}), we obtain the Einstein equation as
\begin{equation}
G_{\mu \nu }+\Lambda g_{\mu \nu }+m^{2}\chi _{\mu \nu }=0,
\label{3}
\end{equation}
where the Einstein tensor $G_{\mu \nu}$ and massive term $\chi_{\mu \nu}$ have the following forms
\begin{eqnarray}
G_{\mu \nu }&=&R_{\mu \nu }-\frac{1}{2}g_{\mu \nu }R,\\
\chi _{\mu \nu } &=&-\frac{c_{1}}{2}\left( \mathcal{U}_{1}g_{\mu \nu }-%
\mathcal{K}_{\mu \nu }\right) -\frac{c_{2}}{2}\left( \mathcal{U}_{2}g_{\mu
\nu }-2\mathcal{U}_{1}\mathcal{K}_{\mu \nu }+2\mathcal{K}_{\mu \nu
}^{2}\right) -\frac{c_{3}}{2}(\mathcal{U}_{3}g_{\mu \nu }-3\mathcal{U}_{2}%
\mathcal{K}_{\mu \nu }+  \notag \\
&&6\mathcal{U}_{1}\mathcal{K}_{\mu \nu }^{2}-6\mathcal{K}_{\mu \nu }^{3})-%
\frac{c_{4}}{2}(\mathcal{U}_{4}g_{\mu \nu }-4\mathcal{U}_{3}\mathcal{K}_{\mu
\nu }+12\mathcal{U}_{2}\mathcal{K}_{\mu \nu }^{2}-24\mathcal{U}_{1}\mathcal{K%
}_{\mu \nu }^{3}+24\mathcal{K}_{\mu \nu }^{4}).  \label{5}
\end{eqnarray}
In order to solve the equation, we take the ansatz metric
\begin{equation}
ds^{2}=-f(r)dt^{2}+f^{-1}(r)dr^{2}+r^{2}d\phi ^{2},
\label{6}
\end{equation}
where $f(r)$ is an arbitrary function of radial coordinate. Then after choosing the reference metric \cite{Vegh:2013sk}
\begin{equation}
f_{\mu \nu}=diag(0,0,c^{2}h_{ij}),
\label{7}
\end{equation}
where $c$ is the positive constant, we can reduce $\mathcal{U}$'s as
\begin{equation}
\mathcal{U}_{1}=\frac{c}{r},\;\;\;\;\;\mathcal{U}_{2}=\mathcal{U}_{3}=\mathcal{U}_{4}=0,
\label{8}
\end{equation}
which means that the only contribution of massive terms come from $\mathcal{U}_{1}$  in 3 dimensional case.

To proceed, we work with $L=1$ so that $\Lambda=-1$, then the independent Einstein equations are
\begin{eqnarray}
&&rf^{\prime }(r)-2r^{2} -m^{2}cc_{1}r =0, \label{9} \\
&&\frac{r^{2}}{2}f^{\prime \prime }(r)-r^{2} =0,
\label{10}
\end{eqnarray}
which correspond to the $tt$ (or $rr$) and $\phi \phi$ components of equation (\ref{3}), respectively. The metric function can be solved from the equation of motion (\ref{10}) as 
\begin{equation}
f(r)=r^{2}-M+m^{2}cc_{1}r,
\label{11}
\end{equation}
where $M$ is an integration constant related with the total mass of the black hole. Note that in the absence of massive term with $m=0$ , the above solution recovers that for BTZ black hole.

Then, the location of horizon solving from $f(r)=0$ are
\begin{equation}
r_{\pm}=\frac{-m^{2}\pm \sqrt{m^{4}+4M}}{2},
\label{12}
\end{equation} 
where we have set $cc_{1}=1$ for convenience. It is noticed that from  (\ref{11}), one can set $\mid cc_1\mid=1$ without loss of generality, but here we first consider the case with $cc_{1}=1$ to show our main process. Later, we will directly give the corresponding result for the case with $cc_{1}=-1$. $r_{+}$ denotes the event horizon, so that we can define near-horizon region of spacetime by another variable, $\rho$, as
\begin{equation}
r=\frac{-m^{2}+ \sqrt{m^{4}+4M}}{2}(1+\rho)  \; , \; 0< \rho \ll 1.
\label{13}
\end{equation}
Therefore,  the metric (\ref{6}) in the leading order  becomes
\begin{eqnarray}
d\tilde{s}^{2}&= &\tilde{g}_{\mu \nu}dx^{\mu}dx^{\nu} \notag
\\ &=&-\frac{(m^{4}-m^{2}\sqrt{m^{4}+4M}+4M)}{2} \rho dt^{2}+
\frac{(m^{4}-m^{2}\sqrt{m^{4}+4M}+2M)}{(m^{4}-m^{2}\sqrt{m^{4}+4M}+4M)\rho}d\rho^{2}+r_{+}^{2}d\phi^{2}.
\label{14}
\end{eqnarray}
which is the near-horizon metric of massive BTZ black hole.

\section{Horizon String and Complexity growth of tensionless string}\label{Horizon StringC}

In this section, we firstly follow the analysis of \cite{Bagchi:2021ban} and show that the tensionless string will emerge as we put the string near the horizon of this neutral maassive BTZ black hole. Then we construct the action of a tensionless bosonic string in the massive BTZ black hole and derive the complexity growth from the Nambu-Goto action.

We begin with the metric as we have seen in(\ref{6}):
\begin{equation}
ds^{2}=-(r^{2}-M+m^{2}r)\;dt^{2}+\frac{dr^{2}}{(r^{2}-M+m^{2}r)}+r^{2}d\phi^{2}
\label{15}.
\end{equation}
Then we introduce the tortoise coordinates
\begin{equation}
r_{*}=-\frac{2\operatorname{arctanh}[\frac{m^{2}+2r}{\sqrt{m^{4}+4M}}]}{\sqrt{m^{4}+4M}}
\label{16}.
\end{equation}
We can rewrite the metric (\ref{15}) as
\begin{equation}
ds_{*}^{2}=-(r^{2}-M+m^{2}r)(dt^{2}+dr_{*}^{2})+r^{2}(r_{*})d\phi^{2}
\label{17}.
\end{equation}
Combining (\ref{13}) and (\ref{16}) , we could have 
\begin{equation}
\rho \approx \frac{\sqrt{m^{4}+4M} -\sqrt{m^{4}+4M} \tanh{[\frac{1}{2}\sqrt{m^{4}+4M}r_{*}} ]}{m^{2}-\sqrt{m^{4}+4M}} \;,\; r_{*}\rightarrow -\infty
\label{18}.
\end{equation}
Also, we introduce a critical parameter $r_{c}$, when $r_{c} \rightarrow-\infty$ , we have near horizon form
\begin{equation}
r_{*}=\frac{-m^{2}+\sqrt{m^{4}+4M}}{2}(r_{c}+\delta r)
\label{19}.
\end{equation}

Now we consider a string in this background geometry. The Polyakov action is 
\begin{equation}
S=-\frac{\mathcal{T}_{0}}{2}\int d\xi^{2}\sqrt{-h} h^{ab}\partial_{a}X^{\mu}\partial_{b}X^{\nu}g^{*}_{\mu \nu}
\label{20},
\end{equation}
where $g^{*}_{\mu \nu}$ is the metric of (\ref {17}) and the induced metric of worldsheet can be introduced by
\begin{equation}
\begin{aligned}
G_{ab}&=\partial_{a}X^{\mu}\partial_{b}X^{\nu}g^{*}_{\mu \nu} \\&
       =-(r^{2}-M+m^{2}r)(-\partial_{a}t\partial_{b}t+\partial_{a}r^{*}\partial_{b}r^{*})+r^{2}(r_{*})\partial_{a}\phi \partial_{b}\phi.
\end{aligned}
\label{21}
\end{equation}
Finally, in the near-horizon limit, the (\ref{21}) approximates to
\begin{equation}
\begin{aligned}
G_{ab} &\approx \frac{1}{4}(m^{4}+4M)Sech^{2}[\frac{1}{2}(-m^{4}-4M+m^{2}\sqrt{m^{4}+4M})(r_{c}+\delta r)] \times \\&
[-\partial_{a} t\partial_{b}t+\frac{(-m^{2}+\sqrt{m^{4}+4M})^{2}}{4}\partial_{a}(\delta r)\partial_{b}(\delta r)]+ \\&
\frac{1}{4}(m^{2}+\sqrt{m^{4}+4M}\tanh[\frac{1}{2}(-m^{4}-4M+m^{2}\sqrt{m^{4}+4M})(r_{c}+\delta r)] )^{2}\partial_{a}\phi \partial_{b}\phi.
\end{aligned}
\label{22}
\end{equation}
When we plug this into the (\ref{20}), we get the rescaled string tension $\mathcal{T}$
\begin{equation}
\mathcal{T}=\frac{\mathcal{T}_{0}}{\cosh^{2}[-m^{4}-4M+m^{2}\sqrt{m^{4}+4M})(r_{c}+\delta r)]}.
\label{23}
\end{equation}

As we can see, at the near horizon limit $r_{c}\rightarrow-\infty$, the string tension will be gone and null string will emerge.

Next, we will construct the action of a tensionless bosonic string in the massive BTZ black hole. It is known that the Nambu-Goto action for usual bosonic string with tension $\mathcal{T}$ is 
\begin{equation}
S=\mathcal{T} \int d\tau d\sigma \sqrt{-det\;h_{ab}},
\label{24}
\end{equation}
where $(\tau,\sigma)$ are the worldsheet parameters of the string, and  $h_{ab}$ is the induced metric
\begin{equation}
h_{ab}=\partial_{a} X^{\mu}\partial_{b}X^{\nu} \eta_{\mu \nu},
\label{25}
\end{equation}
with $X^{\mu}$ the spacetime coordinates of string, and $\eta_{\mu \nu}$ the flat metric. Since in our study, we will focus on the string that stays around the near-horizon geometry (\ref{14}), so the string's tension will disappear, as we have seen in (\ref{23}). Subsequently, the  Nambu-Goto action (\ref{24}) for the horizon string, i.e., $\mathcal{T} \rightarrow0$, takes the form \cite{Schild:1976vq,Isberg:1993av,Bagchi:2013bga}
\begin{equation}
S=\int d\tau d\sigma V^{a}V^{b}\partial_{a} X^{\mu}\partial_{b}X^{\nu} \tilde{g}_{\mu \nu},
\label{26}
\end{equation}
where $\tilde{g}_{\mu \nu}$ is our target metric with expression (\ref{14}), and $V^{a}$ is vector density which will be fixed as $V^{a}=(1,0)$ with certain gauge transformation.
To proceed, we assume that the horizon string moves in the subspace, so for worldsheet coordinates, we take $t=\tau$ , $\rho=\sigma$, and perform $X^{\mu}(\tau,\sigma)$ to be
\begin{eqnarray}
X^{\mu}(\tau,\sigma)= \left( \begin{matrix} &t\rightarrow\tau \\& \rho\rightarrow\sigma  \\& \phi\rightarrow v\tau+\xi(\sigma)
\end{matrix}  \right),
\label{27}
\end{eqnarray}
where  $v$ denotes the velocity of a string relative to the black hole and $\xi(\sigma)$ is a function which determines the shape of string. Then the components of induced metric (\ref{25}) are
\begin{equation}
\begin{aligned}
&h_{\tau \tau}=-\frac{1}{2}(m^{4}-m^{2}\sqrt{m^{4}+4M}+4M)\rho +v^{2}\frac{(-m^{2}+\sqrt{m^{4}+4M})^{2}}{4} , \\&
h_{\tau \sigma}=h_{\sigma \tau}=v\xi^{'}(\sigma)\frac{-m^{2}+\sqrt{(m^{4}+4m})^{2}}{4}, \\&
h_{\sigma \sigma}=\frac{(m^{4}-m^{2}\sqrt{m^{4}+4M}+2M)}{(m^{4}-m^{2}\sqrt{m^{4}+4M}+4M)\rho}+[\xi '(\sigma)]\frac{(-m^{2}+\sqrt{m^{4}+4M})}{4}.
\end{aligned}
\label{28}
\end{equation}
Then the action growth of the tensionless string is reduced as 
\begin{equation}
\begin{aligned}
\frac{dS}{dt} \big\mid_{cc_1=1} &=\int_{0}^{\epsilon} d\rho \;h_{\tau \tau}=\int_{0}^{\epsilon} d\rho \;(\partial_{\tau}X^{t} \partial_{\tau}X^{t}X^{t}\tilde{g}_{tt}+ \partial_{\tau}X^{\phi} \partial_{\tau}X^{t}X^{\phi}\tilde{g}_{\phi \phi})\\&
  = \frac{(- m^{2}+\sqrt{m^{4}+4M})^{2}}{4}v^{2}\epsilon-\epsilon^{2}(\frac{m^{4}}{4}+M - \frac{m^{2}\sqrt{m^{4}+4M}}{4}),
\end{aligned}
\label{29}
\end{equation}
which corresponds to the complexity growth in terms of the CA conjecture  for $cc_1 = 1$. With the same procedure, we then obtain the complexity growth for $cc_1 = -1$ as
\begin{equation}
\frac{dS}{dt} \big\mid_{cc_1=-1}=\frac{(m^{2}+\sqrt{m^{4}+4M})^{2}}{4}v^{2}\epsilon-\epsilon^{2}(\frac{m^{4}}{4}+M +\frac{m^{2}\sqrt{m^{4}+4M}}{4}).
\label{30}
\end{equation}

 We should notice here that the difference between (\ref{29}) and (\ref{30}) exists in the signature in front of  $m^{2}$.  Here  $\epsilon$ should satisfy $\epsilon\ll 1$ since the string is very close to the horizon. In the following, we shall analyze the effect of string velocity, black hole mass and graviton mass on this action growth and the choice of $cc_{1}$.

\subsection{Properties}
Firstly, we turn off the graviton mass with $m=0$, and study the effect of the BTZ black hole mass on the tensionless string action. In this case, the action growth \eqref{29} and \eqref{30} has a simple and same form
\begin{equation}
\frac{dS}{dt}=M(v^{2}\epsilon -\epsilon^{2})
\label{31}.
\end{equation}
We can see that with fixed black hole mass $M$ and $\epsilon$, the action growth is a quadratic function of string velocity, implying that the action growth has a minimum value when the string is at rest. And this minimal value should tend to zero because the $\epsilon$ is  very small. As the string moves faster, this action growth increases till it reaches  maximum value when the string velocity approaches the speed of light. This phenomenon is different from those found in the cases that the string are on the boundary \cite{Nagasaki:2018csh,Zhou:2021vsm}, in which the action growth has a maximum value when the string is at rest while  it tends to zero in certain situations when the string's velocity approaches zero. Moreover, it is obvious from \eqref{31} that  with fixed velocity and $\epsilon$, the action growth behaves as a monotonically increasing function of  the black hole mass, similar to the observations for string on the boundary in \cite{Nagasaki:2017kqe}. The possible explanation for these novel phenomena will be given later soon.

\begin{figure}[ht!]
 \centering
  \includegraphics[width=7cm]{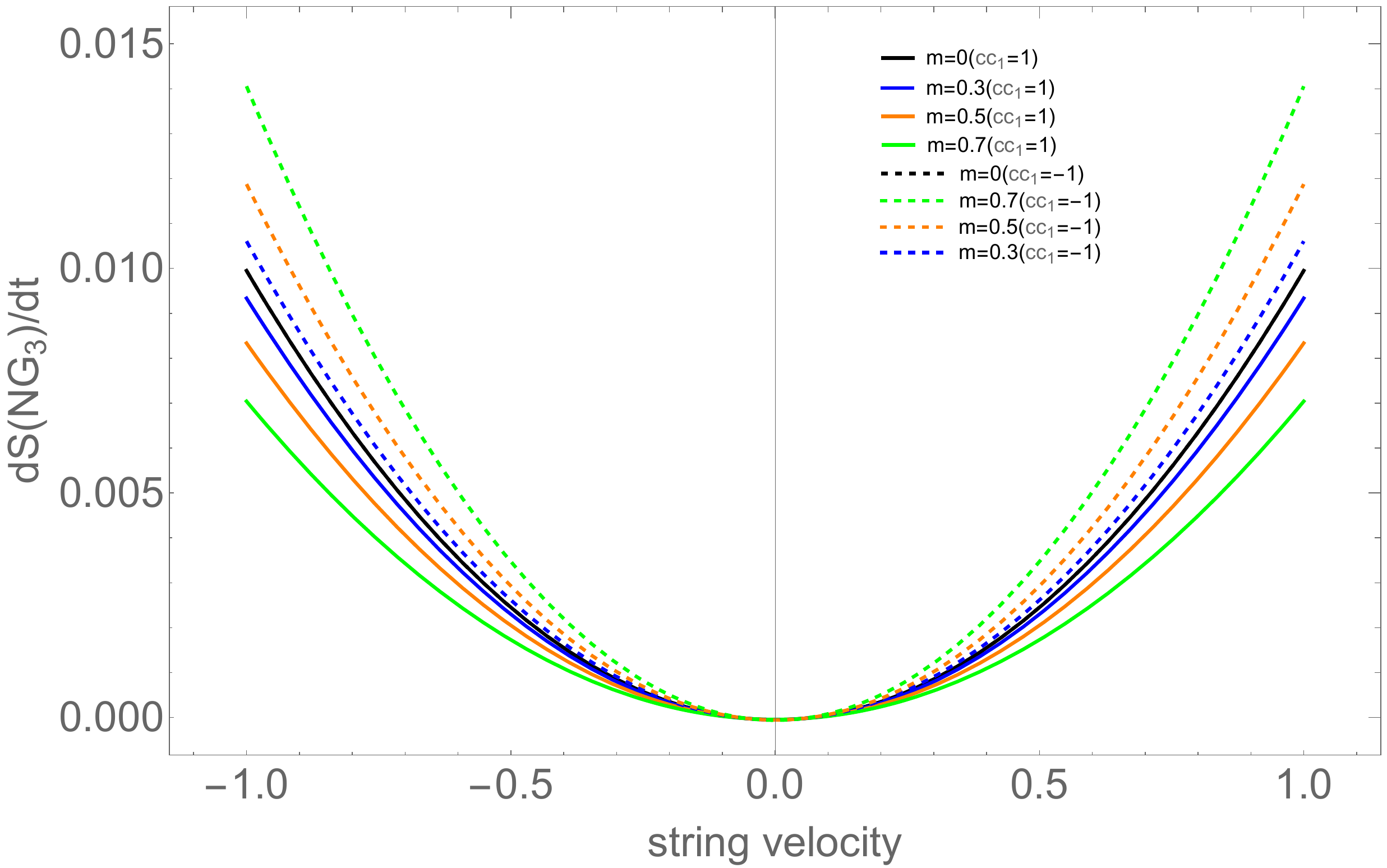}\ \hspace{0.1cm}
     \includegraphics[width=7cm]{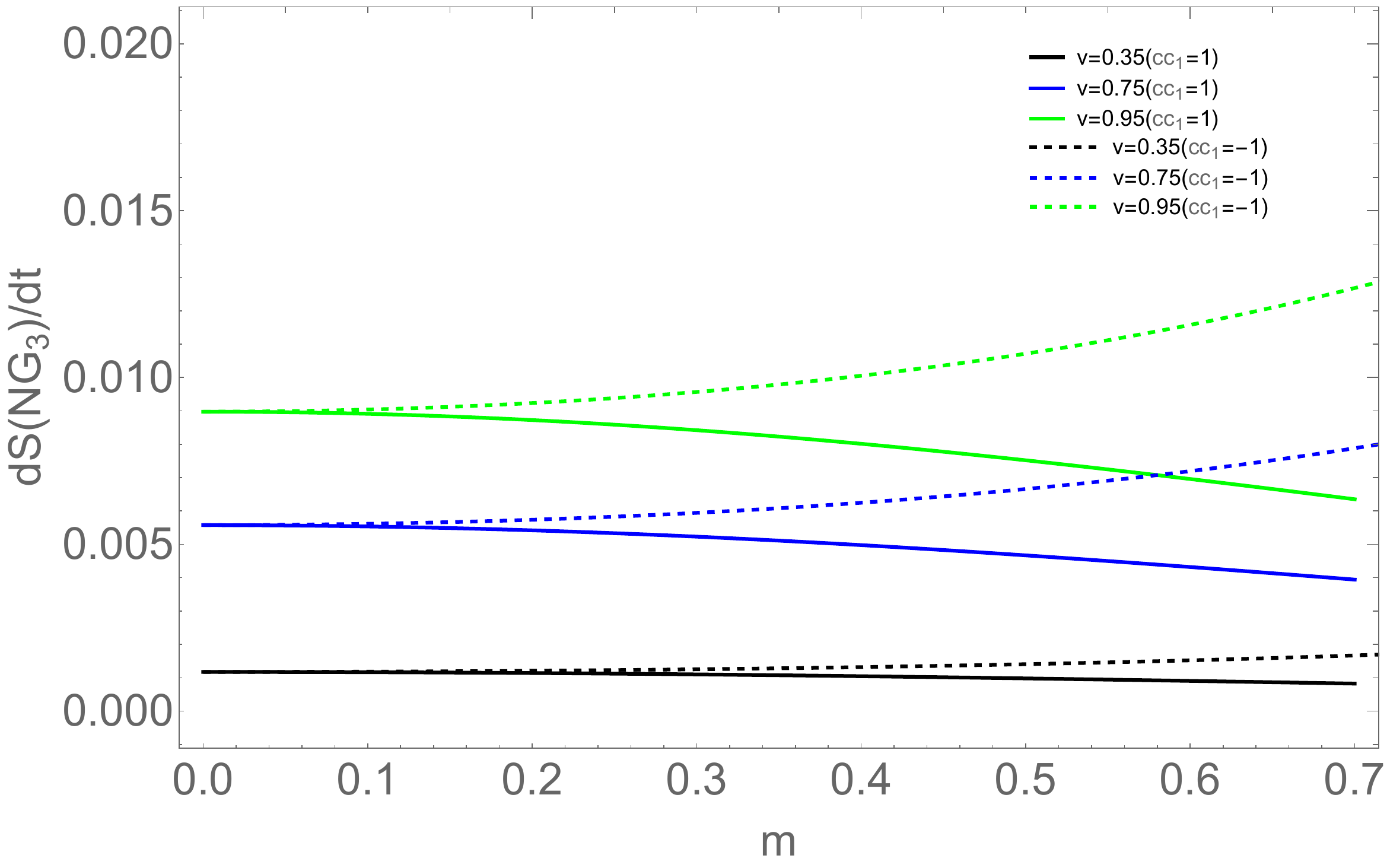}\ \hspace{0.1cm}
    \caption{Left: action growth v.s string velocity in massive BTZ black hole for different graviton masses $m$. Right: action growth v.s graviton mass in massive BTZ black hole for different string velocities. The dashed curves denote the choice of $cc_{1}= -1$ and the solid curves denote the choice of $cc_{1}= 1$. Here we set the black hole mass $M=2$.}
 \label{fig:M2mV}
\end{figure}

Then, we turn on the graviton mass and study the effect of $m$ on the action growth, and we show the results of two choices of $cc_{1}$ with fixed $M = 2$ in Fig.\ref{fig:M2mV}. In the left plot, we show the relation between action growth and string velocity for massive black hole. Here the dashed curves denote the choice of $cc_{1}= -1$ and the solid curves denote the choice of $cc_{1}= 1$. In both cases, the minimal growth appears when the string is stationary, and as the string moves faster, the action growth becomes larger, which is similar to that in BTZ case with  $m=0$. Moreover, the overall values of action growth of $cc_{1}=-1$ are larger than the choice of $cc_{1}=1$. This is reasonable because the radius of  event horizon for $cc_{1}=-1$ is $r_+=\frac{m^{2}+\sqrt{m^{4}+4M}}{2}$, which is always larger than \eqref{12} for $cc_{1}=1$, so the nearly horizon string could be closer to boundary.
In addition, a subtle difference is that  as the graviton mass increases, the action growth for $cc_{1}=-1$ increases while it decreases for $cc_{1}=1$, which are explicitly shown in the right plot. This feature can also be easily read off from \eqref{29} and \eqref{30}. 

%%%%%%%%%%%
%%%%%%%%%%%

\subsection{Discussions}
Next, we shall give possible explanation for why the near-horizon limits of string are very different from the tensile string at the boundary. This could stem from two aspects.
\begin{itemize}
\item According to the expression \eqref{24}, we know that the tension actually contributes to the action growth. Keeping this in mind, we separate \eqref{29} and \eqref{30} into two parts
\begin{equation}
\begin{aligned}
\frac{dS}{dt}&=\int_{0}^{\epsilon} d\rho \;(\partial_{\tau}X^{t} \partial_{\tau}X^{t}X^{t}\tilde{g}_{tt}+ \partial_{\tau}X^{\phi} \partial_{\tau}X^{t}X^{\phi}\tilde{g}_{\phi \phi})\\&
  = \mathcal{T}_{1}\int_{0}^{\epsilon} d\sigma-\mathcal{T}_{2}\int_{0}^{\epsilon}d\sigma\rho,
\end{aligned}
\label{32}
\end{equation}
where $\mathcal{T}_{1}$ and $\mathcal{T}_{2}$ are two effective tensions with the formulas
\begin{equation}
\mathcal{T}_{1} \mid_{cc_1=\pm 1}=v^{2}\frac{(\mp m^{2}+\sqrt{m^{4}+4M})^{2}}{4} ,\; \mathcal{T}_{2} \mid_{cc_1=\pm 1}=\frac{1}{2}(m^{4} \mp m^{2}\sqrt{m^{4}+4M}+4M).
\label{33}
\end{equation}
{These two effective tensions are functions of  the string's velocity, black hole mass, and graviton mass. When the string is at rest, the tension $\mathcal{T}_{1}$ vanishes or the string becomes very ``floppy", while the second term of \eqref{32} should be insignificant because of the near-horizon limit. However, when the string is moving with an increased velocity, this ``floppy" string will strengthen and contributes to the action growth.}

  \item In the case of the tensile string, one usual assumes that the tension of the string is constant, either the string is moving or stationary.  As a result, when the string remains stationary, it takes the maximum value of action growth, and as the velocity increases, the action growth decrease to zero when the string approaches the speed of light \cite{Nagasaki:2017kqe, Nagasaki:2021ldz, Nagasaki:2018csh, Zhou:2021vsm}. It also means that the effective tension should change at the speed of light, which may stretch the string to infinity and get ``floppy,"  and then reduce to the near-horizon limit case.
  \end{itemize}

\section{Closing remarks}\label{summary}
In this work, we studied the complexity growth with a  tensionless probe string in the near-horizon limit of  massive BTZ black hole. We focused on the Nambu-Goto action growth and employed the CA conjecture to explore the effect of tensionless string on the complexity growth with momentum relaxation. We found  novel features of the complexity growth comparing to that in the case with the probe string in the boundary. The results are summarized as follows.

In BTZ case, the minimal complexity growth appears when the string is motionless, and as the string moves faster, the action growth becomes larger until its velocity approaches the speed of light. This phenomenon is very different from that for the tensile string. It is argued that when the string begins to move with faster velocity, the ``floppy" string will gain strength and contribute to the action growth. The action growth increases monotonically with increasing black hole mass at all chosen velocities. 
%This could mean that a larger system contains more information and is thus more complex.
In massive BTZ case, we compared the two choices of $cc_{1}$. We found that in both cases, the minimal action growth appears when the string is stationary and also increases monotonically with velocities. But the overall values of action growth of $cc_1 = -1$ are larger than the choice of $cc_1 = 1$. We think the reason for this is that the radius of the event horizon for $cc_1=-1$ is bigger than the \eqref{12} for $cc_1=1$, so the nearly horizon string could be closer to the boundary(for more inspired discussion, see below). In addition, a subtle difference is that as the graviton mass increases, the action growth for $cc_1 =-1$ increases. As opposed to that, for $cc_1 =1$, the value of action growth
decreases as the graviton mass increases.

Our studies show that the overall values of complexity growth in the tensionless case
are much smaller than that in the tensile case, which could arise some enlightening discussion. Firstly, following the holographic idea of renormalization group flow theory, this finding means that the complexity growth at the IR side (near-horizon limit string) is smaller than that at the UV side (boundary tensile string), which could imply that there may exist certain ``complexity c-function", at least in two dimensions as that for entanglement entropy \cite{Nishioka:2018khk}.   Secondly, inspired by the above discussion, we may define a flow, namely ``complexity flow" , which depicts the changes in complexity along a certain trajectory. Then with this definition,  one can reconstruct the complexity from the boundary by using the ``complexity flow". These  would be interesting directions and deserve further exploration, which definitely will shed some light on the understanding of complexity in quantum field theory and holography.

\acknowledgments{}
We appreciate Arjun Bagchi, M. M. Sheikh-Jabbari and Daniel Grumiller for helpful corresponding.  Moreover, we are grateful to Profs. Ya-Peng Hu and Xiao-Mei Kuang for valuable discussions. This work is supported by National Natural Science Foundation of China (NSFC) under grant Nos. 12247170, 12175105, 11575083, 11565017, 12147175, Top-notch Academic Programs Project of Jiangsu
Higher Education Institutions (TAPP).

\bibliographystyle{style2}
\bibliography{Ref}
\end{document}